# Signal and Noise scaling factors in digital holography under weak illumination conditions: relationship with shot-noise

M. Lesaffre[1], N. Verrier[2], M. Atlan[1], M. Gross*[2]
[1]*Institut Langevin - UMR 7587 CNRS-ESPCI Paristech
1, rue Jussieu 75005 Paris ;* [2]*Laboratoire Charles Coulomb - UMR 5221 CNRS-UM2 Université Montpellier II Place Eugène Bataillon 34095 Montpellier cedex*
*gross@lkb.ens.fr

**Abstract:** An experimental study on how reconstructed image signal and noise scale with the holographic acquisition and reconstruction parameters is proposed. Comparison with Monte-carlo simulations is performed to emphasize that the measured noise is the shot-noise.
**OCIS codes:** (090.0090) Holography; (090.1995) Digital Holography; (120.2880) Holographic Interferometry

## 1. Introduction

Holographic signal $EE_{LO}^*$, results in the interference between the object field $E$ and a reference field $E_{LO}$, whose amplitude is generally higher than that of the object field. Heterodyne gain ($EE_{LO}^* \gg E^2$) is therefore obtained, making holographic detection highly suitable for the measurement of weak object fields.

Working with off-axis heterodyne holography also makes it possible to shift frequency of the signal in both time and space. Therefore, a high efficiency filtering of holographic signal in both time and space domains is possible. Due to all these aspects, the digital off-axis heterodyne holography operates with an ultimate sensitivity [1]. Various domains such as, *in-vivo* Doppler imaging of human breast [2], localization and tracking of metallic nano-beads [3, 4], and small vibration amplitude sideband analysis [5-7], benefit from heterodyne holography properties.

As far as, when working with weak signal, the limiting noise is the shot-noise, knowledge of its scaling law, according to the recording and reconstruction parameters is important [8-9].

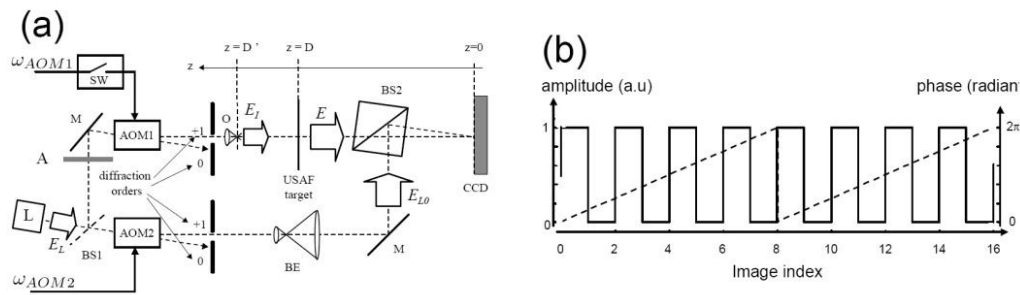

**Fig. 1** (a) Experimental set-up. L: Laser diode; BS1, BS2: Beamsplitters; AOM1, AOM2: Acouto-optic modulators; O: Microscope objective; BE: Beam expander; M: mirror; A: Neutral density. (b) Intensity (solid) and phase (dash) of the USAF illumination beam.

## 2. Experimental set-up and reconstruction procedure

The experimental set-up, proposed Fig. 1(a), is a classical heterodyne holography arrangement [10]. However, a switch has been added on the generator driving AOM1. It enables, as illustrated by Fig 1(b), to switch on/off the

USAF illumination beam $E_l$ from one image to the other. Moreover, the reference beam phase-shift is adjusted so that to perform 8-phases phase-shifting: $\omega_{AOM1} - \omega_{AOM2} = \omega_{CCD}/8$, where $\omega_{CCD}$ is the CCD sensor acquisition rate. This experimental configuration permits to obtain various image combinations from the same acquisition run. For instance, using only the even, or odd frames enables to obtain, 4-phases phase-shifting holograms with 4, 8, 16, or 32 frames of illuminated ($H_4$, $H_8$, $H_{16}$, $H_{32}$), or un-illuminated ($H'_4$, $H'_8$, $H'_{16}$, $H'_{32}$) object. Working using one frame out of four can be used for two-phase illuminated ($H_2$), or un-illuminated ($H'_2$) holograms. Moreover, by calculating the difference of two consecutive holograms, one can obtain a zero order grating term-free hologram ($H_{1,1}$) [11].

Holographic reconstruction, illustrated Fig. 2, is realized over 1024x1024 pixels using angular spectrum propagation [12], with spatial filtering of the unwanted diffraction terms over 400x400 pixels [13].

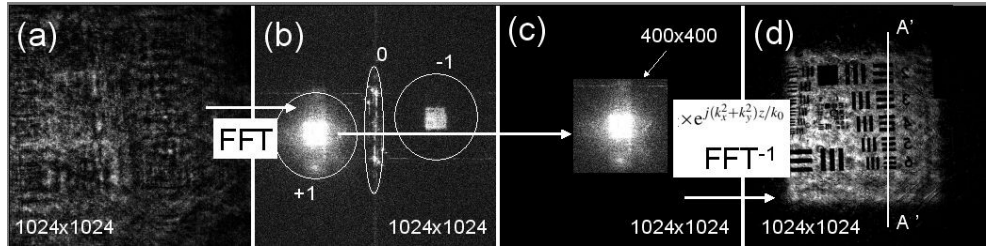

**Fig. 2** (a) Holographic signal intensity $|H_4(x, y, 0)|^2$ in the CCD plane; (b, c) Holographic signal in Fourier space without (b) and with (c) 400x400 pixels spatial filtering ; (d) Reconstructed image of the USAF target $|H_4(x, y, z = D)|^2$.

## 3. Experimental results

To investigate the influence of the recording parameter on the signal and noise, we acquired and reconstructed, under weak illumination, holograms for various number of frames, either illuminated or not ($H_{1,1}$, $H_2$, $H'_2$, ...,$H_{32}$, $H'_{32}$), and for different spatial filter areas (from 100x100 to 400x400 with a step of 2 in area).

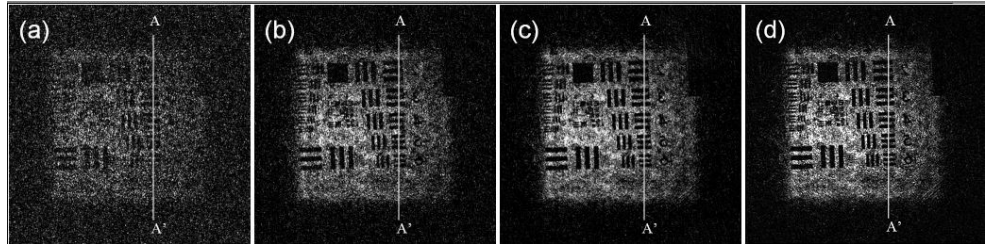

**Fig. 3** USAF target reconstructed images $|H(x, y, D)|^2$ for various $N_b$ frames : (a) $N_b$=2 frames: $|H_2|^2$, (b) 4 frames : $|H_4|^2$, (c) 8 frames : $|H_8|^2$ et (d) 16 frames : $|H_{16}|^2$. Saptial filter 282 x 282 pixels.

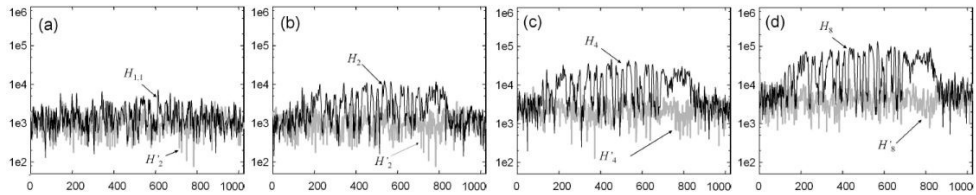

**Fig. 4** Intensity profiles of $|H(x, y, D)|^2$ along AA ' line of Fig.3. Dark curves with illumination: $H_{1,1}$ (a), $H_2$ (b), $H_4$ (c) et $H_8$ (d), light curves without: $H'_2$ (a), $H'_2$ (b), $H'_4$ (c) et $H'_8$ . In $e^2$ per pixel (e: photo electron).

Obtained results are depicted Figs. 3 and 4. It is here easily noticeable that the more the image number, the higher the signal. This aspect is more quantitavely illustrated by Fig. 4. Here, intensity profiles of $|H(x, y, D)|^2$ along AA ' line, are plotted with (dark curve), and without (light curve) illumination. The latter case corresponds to the recorded noise. One can notice, from Fig. 4 that the signal is proportional to $N_b^2$.

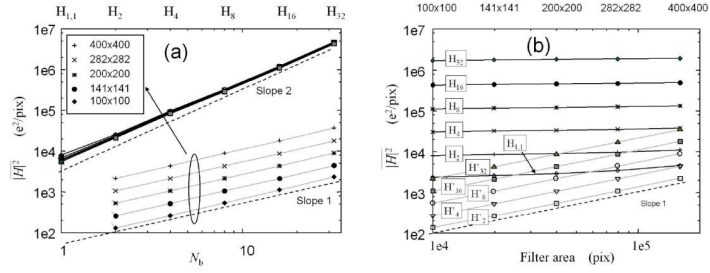

**Fig. 5** (a) Mean intensity $|H|^2$ versus the number of frame $N_b$ with illumination: $N_b = 1$ for $H_{1,1}$, $N_b = 2$ pour $H_2$ or $H'_2$, ..., and $N_b = 32$ for $H_{32}$ or $H'_{32}$, (b) as a function of the spatial filter area. Dark curves with illumination, light curves without illumination.

In order to be more quantitative, the mean intensity of the recorded signal is calculated for different numbers of frames (from 1 to 32), and for different spatial filter areas (100x100 to 400x400 pixels). Results are proposed Fig. 5. As previously stated, the signal is proportional to $N_b^2$, and does not depend on the area of the spatial filter. The background noise, obtained from the un-illuminated frames is proportional to $N_b$ and to the spatial filter area. Therefore, the signal to noise ratio (SNR) is proportional to the acquisition duration as theoretically predicted for an ideal coherent detection. Moreover, the SNR is proportional to the inverse of the filter area. Nevertheless, the area of the filter should not be dramatically reduced is order to maintain reconstructed image resolution.

## 4. Comparison with shot-noise

As shot-noise is a Poissonian noise, Monte-Carlo method is suited for its simulation. Taking a reference frame, and adding a random Gaussian noise to each pixel makes it possible to obtain a first shot-noised frame. This operation is repeated to obtain 64 Monte-Carlo frames. Then, Monte-Carlo holograms ($H''_2, H''_4, ..., H''_{32}$) are constructed as the previously acquired holograms with ($H_2, H_4, ..., H_{32}$) or without ($H'_2, H'_4, ..., H'_{32}$) illumination. Therefore, obtained Monte-Carlo holograms only contain noise, which is the minimal theoretical noise awaited in an ideal holographic detection.

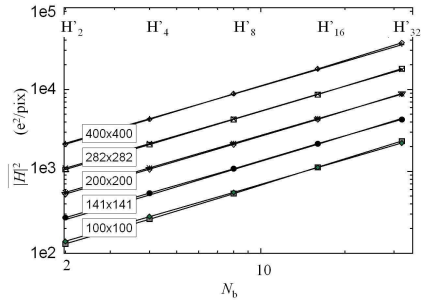

**Fig. 6** Mean reconstructed intensity $|H|^2$ versus number trames $N_b$: $N_b = 2$ for $H'_2$ or $H''_2$, $N_b = 4$ for $H'_4$ or $H''_4$..., and $N_b = 32$ for $H'_{32}$ or $H''_{32}$. Frames are recorded without illumination $H'_x$ or Monte Carlo simulated $H''_x$. Filter area ranges from 100x100 to 400x400 pixels.

To prove that our experimental noise is the shot-noise, mean reconstructed intensities without illumination ($H'_n$) and simulated intensities ($H''_n$) are compared for spatial filter areas ranging from 100x100 to 400x400 pixels. Results plotted Fig. 6 reveals the good agreement between experimental and simulation data. Thus, the noise of our holographic detection is the shot-noise.

## 5. Conclusion

By recording, under weak illumination conditions, holograms of a USAF target, we obtain for both the signal and the noise, scaling laws of a shot-noise limited detection. The intensity signal of the reconstructed image $|H|^2$, is proportional to the squared acquisition duration, whereas the noise, which is equal to the shot noise [8], is proportional to the acquisition duration, and to the number of optical modes used for reconstruction (area of the spatial filter in pixel). These results validate our previous studies in digital holography with weak illumination [1-6].